# The different shapes of spin textures as a journey through Brillouin zone chiral and polar symmetries: application to spin-valleytronics

Carlos Mera Acosta[1,2], Linding Yuan[1], Gustavo M. Dalpian[2], and Alex Zunger[1]

[1]*Renewable and sustainable Energy Institute, University of Colorado, Boulder, CO 80309, USA*
[2]*Center for Natural and Human Sciences, Federal University of ABC, Santo Andre, SP, Brazil*

Crystallographic space group symmetry (CPGS) such as polar and nonpolar crystal classes have long been known to classify compounds that have spin-orbit-induced spin splitting. While taking a journey through the Brillouin Zone (BZ) from one $k$-point to another for a fixed CPGS, it is expected that the wavevector point group symmetry (WPGS) can change, and consequently a qualitative change in the *texture* of the spin polarization (the expectation value of spin operator $\vec{S}_{nk_0}$ in Bloch state $u(n,k)$ and the wavevector $k_0$). However, the nature of the *spin texture* (ST) change is generally unsuspected. In this work, we determine a full classification of the linear-in-$k$ spin texture patterns based on the *polarity* and *chirality* reflected in the WPGS at $k_0$. The spin-polarization vector $\vec{S}_{nk_0}$ controlling the ST is bound to be parallel to the rotation axis and perpendicular to the mirror planes and hence, symmetry operation types in WPGSs impose symmetry restriction to the ST. For instance, the ST is always parallel to the wavevector $k$ in non-polar chiral WPGSs since they contain only rotational symmetries. Some consequences of the ST classification based on the symmetry operations in the WPGS include the observation of ST patterns that are unexpected according to the symmetry of the crystal. For example, it is usually established that spin-momentum locking effect (spin vector always perpendicular to the wavevector) requires the crystal inversion symmetry breaking by an asymmetric electric potential. However, we find that polar WPGS can have this effect even in compounds without electric dipoles or external electric fields. We use the determined relation between WPGS and ST as a design principle to select compounds with multiple ST near band edges at different $k$-valleys. Based on high-throughput calculations for 1481 compounds, we find 37 previously fabricated materials with different ST near band edges. The ST classification as well as the predicted compounds with multiple ST can be a platform for potential application for spin-valleytronics and the control of the ST by accessing to different valleys.

## I. Introduction

Whereas *symmetry* generally allows or forbids numerous effects in solid state and molecular science, a recurring question is what aspect of group symmetries is responsible for given type of phenomena. For example, symmetries contained in the Crystallographic Point Group Symmetry (CPGS) establish the enabling conditions for macroscopic properties, such as electric polarization [1], magnetization [2], circular dichroism [3], and pyroelectricity [4]. The CPGS is, however, insufficient to universally describe all materials properties in crystals. In fact, wavevector dependent effects are enabling by other elements of symmetry such as the *wavevector point group symmetry* (WPGS) – the little point group of specific wavevectors $k_0$ in the Brillouin zone (BZ). For instance, symmetry protection of exotic Fermions [5,6] and energy band anti-crossing [7] depends on the WPGS. The Zeeman-type spin splitting (SS) [8,9] is an example of enabling physical mechanisms that seem to contradict the macroscopic crystal symmetry. Specifically, in contrast to the Zeeman effect in magnetic compounds, the Zeeman-type effect is observed in non-magnetic compounds (i.e., CPGS preserving the time-reversal symmetry) but at $k$-point with WPGS breaking the time-reversal (TR) symmetry. Naturally, effects enabled by the CPGS are only allowed at those special wavevectors at which the two symmetries coincide (WPGS=CPGS). However, it is expected that other wavevectors lead to very different effects.

Overlooking the distinction between different physics enabled by WPGS *vs* that enabled by the CPGS has often created an incompleteness of the symmetry classification of spin-related phenomena and their wavevector dependence. A curious historical development in this regard has been the association of the CPGS with the 'texture' of the spin-polarization $\vec{S}_{nk_0}$ – the expectation values of spin operators $\hat{S}_i$ in a given Bloch wavefunction $u(n,k)$ that is centered at a specific $k_0$ with $n$ referring to a Bloch band. Specifically, figure 1 illustrates different shapes of spin texture (ST) that have been observed, including *radial* ST ($\vec{S}_{nk} \parallel \vec{k}$) [10–12], or *tangential* ST ($\vec{S}_{nk_0} \perp \vec{k}$) [13–15], or the *tangential-radial* ST [16,17]. These observations were generally established for highly specific wavevectors $k_0$ that satisfy WPGS=CPGS, e.g., the $\Gamma$ point in GaAs (F$\bar{4}$3m) [16] and the Z point in GeTe (R3m) [13,14]. For this reason, ST shapes were often associated with the presence or absence of *crystallographic* inversion symmetry in the CPGS, rather than with the WPGS of the specific wavevector $k_0$. Furthermore, when different ST shapes were noted in different functional materials, it was tempting to associate the specific ST with the particular underlying functionality. For instance, the observation of *tangential* ST in some ferroelectrics has been associated with the physics of electric polarization [14,18,19]. However, not all ferroelectrics have such ST type [20]. Similarly, the observation of *tangential* ST in some topological insulators has been associated with their topological character; however, normal insulators can also have this very same ST [21].

In $k \cdot p$ effective Hamiltonians $\mathcal{H}(k \to k_0)$, for wavevectors $k$ around the origin $k_0$ of the expansion, the ST is properly determined by the little point group of wavevector $k_0$ [22,23]. However, despite extensive applications of the $k \cdot p$ effective Hamiltonians to study STs for particular wavevectors $k_0$, associations of the resulting ST behavior with the crystallographic CPGS symmetry rather than the *wavevector* symmetry WPGS abound. For example, the *tangential* ST (Fig1.b) seen in bulk compounds is often associated with the Rashba physics of breaking the *crystallographic* inversion symmetry via asymmetric electric potentials (i.e., electric field or bulk electric polarization) [24], rather than with the WPGS of the particular wavevector studied. Although it is properly expected that while taking a journey through the BZ from one point $k_0$ to another, the $k$ vector symmetry and the thus the ST might qualitatively change, the nature of the change is generally unsuspected. This position was clearly expressed in a recent paper studying the *radial* spin texture of Weyl fermions in chiral Tellurium [11] concluding that "A full classification of the spin vector field geometry is beyond the scope of this study, and it will be the subject of a future investigation."

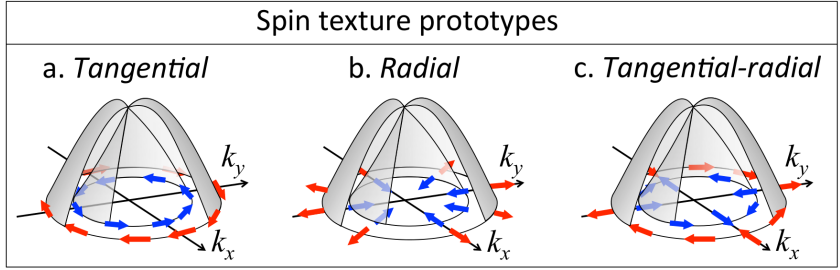

Figure 1: Schematic representation of the (a) *tangential* texture $\vec{S}_{nk} \perp \vec{k}$, (b) *radial* texture $\vec{S}_{nk} \parallel \vec{k}$, and (c) *tangential-radial* texture at a two-dimensional plane $k_{xy}$ normal to a rotation axis $R_i$.

The present paper discusses a direct resolution of the classification problem of STs. We show that the enabling symmetries underlying the ST types are not a reflection of material functionalities, nor are they caused by the presence or absence of polar fields in the CPGS [24]. Instead, ST shapes are caused by a symmetry principle cutting across different material functionalities: the existence *of specific proper (rotations) and improper symmetries (reflections and inversion) in the wavevector point group symmetry (a subgroup of the CPGS)*. Specifically, we show that the spin-polarization vector $\vec{S}_{nk_0}$ controlling the ST is bound to be parallel to the rotation axis and perpendicular to the mirror planes. This imposes a prediction on the ensuing ST patterns according to the *polarity* and *chirality* of the $k_0$ WPGS. For example, for $k_0$ with non-polar chiral WPGS having more than one rotation symmetry and no mirror symmetry we expect $\vec{S}_{nk_0}$ to be parallel to the rotation axis i.e., that $\vec{S}_{nk_0}(k) \parallel \vec{k}$. Here, chiral (non-chiral) point groups have only proper (both proper and improper) symmetries, while polar (non-polar) point groups have one (more than one) rotation axis. *Thus, a journey throughout the Brillouin zone of a fixed compound reveals ST types corresponding to different rotation and reflection symmetries of the little group of $k$.* For instance, we expect (and confirm) that compounds with non-polar crystallographic symmetry (i.e., without electric dipoles or electric fields) can show the *tangential* ST (i.e., Rashba-like ST) at the wavevector $k_0$ whose WPGS is polar. This illustrates that the breaking of the inversion symmetry mediated by and electric dipole or external electric field being a defining feature of Rashba effect is not a necessary condition for the Rashba ST (Fig. 1), contrary to what is generally assumed by the macroscopic CPGS and used to investigate the formation of spin spirals [24–27] for different classes of materials.

Understanding the association of ST with the WPGS (rather than with the wavevector-independent CPGS) could be a productive basis for design of compounds with target ST and its control by accessing different valleys in the BZ. Thus, we explore the potential application to spin-valleytronics of the proposed *journey throughout* the BZ to access multiple ST types in the same compound. The application of our study is based on the *inverse design approach* [28] for the selection of compounds with single and co-functionalities [7,20]. Contrary to the "*direct approach*" based on the calculation for all possible materials candidates, the inverse design aims first the establishment of the physical mechanisms (or *design principles*) behind the target property, i.e., compounds with multiple ST shapes in the BZ. In the second step of this approach, the design principles are used as filters for the screening of compounds from known materials databases, e.g., the aflow-ICSD database [29]. In the first step, for BZs of 3D non-centrosymmetric Bravais lattices, we apply the relations between the ST shapes and the WPGS *polarity* and *chirality*, determining all possible symmetrically allowed WPGS in all 21 non-centrosymmetric CPGS (i.e., 139 crystallographic space groups). Only polar non-chiral, non-polar chiral, and non-polar non-chiral CPGS can have CPSG allowing high-symmetry $k$-points with different STs. Based on the CPGS and WPGS, we select 1481 fabricated compounds and perform high-throughput DFT band structure calculations for them. Focusing on bands structures in which the relative energy at different valleys is smaller than 100 meV and SS larger than 1 meV, we identify 37 compounds with multiple ST shapes. Examples include non-polar chiral $SiO_2$ ($P6_522$), CPGS=$D_6$, that have radial ST at the high-symmetry point A (WPGS=CPGS) and tangential-radial ST at the high-symmetry point H (WPGS=$D_3$). The SS at these $k$-points are 26 and 6 meV, respectively. The proposed classification of the ST based on the WPGS and the selected compounds in the inverse design process are a potential platform for spin-valleytronics applications.

## II. Classic $k \cdot p$ Hamiltonians used to provide the ST type for the special case of WPGS=CPGS

We next illustrate three linear-in-$k$ relativistic Hamiltonians (Rashba, Weyl, and Dresselhaus) set to a specific wavevector $k_0$ with WPGS equaled that of the crystallographic CPGS, which is not representative of other parts of the BZ, as summarized in Table I. Figure 2 shows the DFT calculated STs at different wavevectors $k_0$ with both WPGS=CPGS and WPGS≠CPGS for the representative compounds described in Table I.

Table I. When the wavevector point group symmetry (WPGS) of $k_0$ equals the global crystallographic crystal point group symmetry (CPGS), i.e., $k_0$ with WPGS=CPGS (shown here for GaAs, GeTe, and Te), then ST type has the highest symmetry. But a journey through the BZ shows the more typical cases where $k_0$ has WPGS≠CPGS, leading to many different types of ST.

| Compounds | CPGS | $k_0$ with WPGS=CPGS | ST for $k_0$ with WPGS=CPGS | $k_0$ with WPGS≠CPGS | ST for $k_0$ with WPGS≠CPGS |
|---|---|---|---|---|---|
| GeTe (*R3m*) | $C_{3v}$ | $\Gamma$ ($C_{3v}$), Z($C_{3v}$) | Rashba | L($C_s$) | Dresselhaus-Rashba |
| Te (*P3$_1$21*) | $D_3$ | $\Gamma$ ($T_d$), A($T_d$) | Weyl | M($C_2$) | Undefined |
| GaAs (*F$\bar{4}$3m*) | $T_d$ | $\Gamma$ ($T_d$) | Dresselhaus | L ($C_{3v}$) | Rashba |

In 1984, Bychkov and E. Rashba established [30–32] that "*if a crystal has a single high-symmetry axis (at least threefold)*", i.e., crystals with polar CPGS, spin bands are described by the linear-in-$k$ spin-orbit coupling (SOC) Hamiltonian,

$$\mathcal{H}_R(k \to k_0) = \lambda_R(k_x\sigma_y - k_y\sigma_x),$$

where the z component of the momentum $\vec{k}$ is set along the high-symmetry axis and $\sigma_i$ are the Pauli matrixes. The Hamiltonian $\mathcal{H}_R$ was historically used to study both two-dimensional compounds with perpendicular electric fields and heterojunctions with interfacial electric dipoles. In these systems, only wavevectors $k_0$ with WPGS=CPGS (e.g., $k_0 = \Gamma$) satisfy the Hamiltonian $\mathcal{H}_R$. The diagonalization of $\mathcal{H}_R$ leads to the *tangential* ST (i.e. $\vec{S}_{nk}$ always perpendicular to the momentum ($\vec{S}_{nk_0}(k) \perp \vec{k}$) or equivalently $\vec{S}_{nk_0}(k) = (-k_y, k_x, 0)/|\vec{k}|$), which is usually referred to as Rashba ST or *spin-momentum locking effect*. In the three-dimensional analogue, i.e., the bulk Rashba effect, compounds with polar CPGS (e.g., BiTeI (*R3m*) and GeTe (*R3m*)) [13,15,33], have an intrinsic non-zero electric dipole that effectively plays the role of the interfacial dipole in heterojunctions. As shown in the first line of Table I, in GeTe (*R3m*), the BZ wavevector $k_0 = Z$ has the Rashba-like ST, as illustrated via relativistic DFT calculations in Fig. 2b. This spin-momentum locking effect is also observed in the surface states near the $\Gamma$ point of BZ of three-dimensional topological insulators [34].

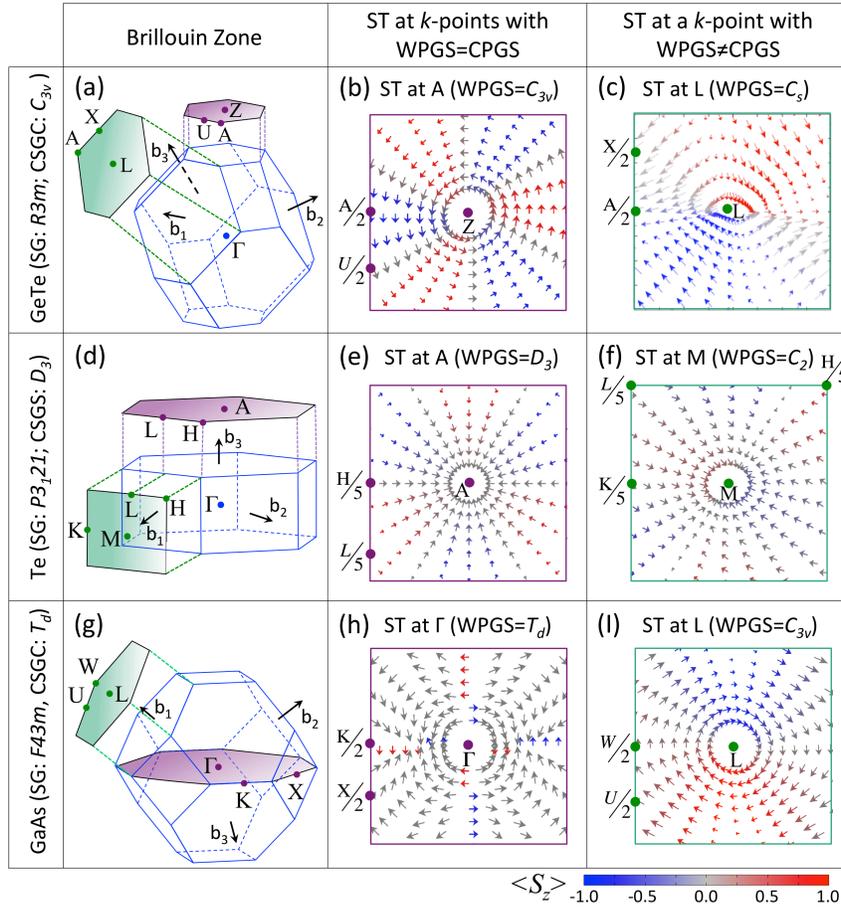

Figure 2. Variations of DFT-calculated Spin Texture (ST) prototypes at different wavevector point group symmetries (WPGS) viewed as a journey through the Brillouin zone of fixed given compounds. We show the Brillouin Zone for (a) GeTe, (d) Tellurium, and (g) GaAs, which are respectively non-polar, polar, and chiral compounds according to their crystallographic point group symmetry (CPGS). The journey departure points for GeTe (b), Te (e), and GaAs (h), are wavevectors $k_0$ that have a point group symmetry equal to the macroscopic CPGS, so the ST are the ones expected from the CPGS, i.e., *tangential* ST for GeTe, *radial* ST for Te, and *tangential-radial* ST for GaAs. However, shifting to $k$-points of lower wavevector symmetries [(c), (f), (i), respectively] shows greatly altered ST shapes for the same compounds.

In crystals with chiral CPGS, at wavevectors $k_0$ with WPGS=CPGS, spin bands are described by the effective Hamiltonian,

$$\mathcal{H}_w(k \to k_0) = \lambda_w(k_x\sigma_x + k_y\sigma_y),$$

which was first proposed by H. Weyl in 1929 [35]. The diagonalization of $\mathcal{H}_w$ results in the *radial* ST (i.e., the spin always parallel to the wavevector, $\vec{S}_{nk_0}(k) \parallel \vec{k}$, or equivalently $\vec{S}_{nk_0}(k) = (k_x, k_y, 0)/|\vec{k}|$), which is historically know as Weyl ST. As shown in the second line of Table I, in the two equivalently chiral enantiomers of bulk Tellurium (*P3₁21* and *P3₂21*) [10–12], at the high-symmetry $k$-points Γ and A (with WPGS=CPGS), the Weyl ST is observed (Fig. 2e). The Hamiltonian $\mathcal{H}_w$ also describe topological Weyl semimetals [36,37].

In compounds with non-polar CPGS, in 1955 G. Dresselhaus determined [38] that for $k_0 = \Gamma$ (WPGS=CPGS), spin bands are described by the Hamiltonian,

$$H_D = \lambda\left[(k_y^2 - k_z^2)k_x J_x + (k_z^2 - k_x^2)k_y J_y + (k_x^2 - k_y^2)k_z J_z\right],$$

where $J_i$ are the components of the total angular momentum operator and $k_z$ is fixed along a rotation axis $R_i$ in the BZ. In the normal plane to $R_i$, the linear-in-$k$ Hamiltonian is given by the Dresselhaus term,

$$\mathcal{H}_D(k \to k_0) = \lambda_D(k_x\sigma_x - k_y\sigma_y).$$

The diagonalization of $\mathcal{H}_D$ leads to the *tangential-radial* ST (i.e., $\vec{S}(n, k_0) = (k_x, -k_y, 0)$), which is usually known as Dresselhaus ST. In the third line of Table I, we present GaAs (*F$\bar{4}$3m*) with non-polar CPGS $T_d$, at $k_0 = \Gamma$ (WPGS=$T_d$), as an historical example of the Dresselhaus ST, as illustrated via quantitative relativistic DFT calculations in Fig. 2h.

Hereafter, we will refer to $\mathcal{H}_R$, $\mathcal{H}_W$, and $\mathcal{H}_D$ as linear-in-$k$ SOC Rashba, Weyl, and Dresselhaus Hamiltonians, respectively. The ST resulting from these Hamiltonians are routinely observed in specific high-symmetry *k*-points with WPGS=CPGS of compounds with non-polar CPGS (such as the Γ point in GaAs (F$\bar{4}$3m) [16] and IrBiSe (P213)) [17]) and compounds with polar CPGS (such as the Z point in GeTe (R3m) [13,14] and BiTeI (P3m1) [15]).

### III.   $k_0$ dependent effects are reflected in the WPGS whereas macroscopic symmetry effects are reflected in the CPGS

A journey through wavevectors in a BZ can visit symmetries different than the macroscopic CGPS. Specifically, for each CPGS, there exists another layer of symmetries of particular wavevectors $k_0$ (show in Table I) in the corresponding BZ [39–41]. This layer of symmetries consist of subgroups of the CPGS and enable specific momentum and band dependent properties in the crystal such as band crossing, anti-crossing [42], topological band inversion [43–45], and topological protection [46]. The inspection to STs at other $k$-points reveals patterns that are not predicted by the CPGS. For instance, for the traditional ferroelectric compound GeTe (*R3m*) with Rashba ST at $k_0$ =Z (polar WPGS and CPGS) [47], Fig. 2c shows the ST obtained from the same relativistic calculation in Fig. 2b, but this time at another wavevector $k_0 = L$ (first line in Table I). The ST reveals a pattern that is not a Rashba-like ST (Fig. 2c). Likewise, in bulk Te, Fig. 2e presents the traditional Weyl ST at $k_0$ =A, which is deformed in another wavevector $k_0 = M$ showing an apparently undefined pattern (Fig. 2f). We will discuss this undefined pattern in the next sections. Finally, for the non-polar compound GaAs with Dresselhaus ST at $k_0 = \Gamma$ (WPGS=CPGS), Fig. 2i shows the typical *Rashba-like* ST obtained quantitatively from the same relativistic band structures used in Fig 2b, but this time at another wavevector $k_0 = L$. The latter example of GaAs illustrates a *Rashba ST in a compound without electric dipoles* (i.e., compound with non-polar CPGS). Additionally, in contrast to the previous suggestion that Rashba ST is attributed to an intrinsic electric dipole and strong atomistic SOC [48], we see that even with relatively weak atomic SOC in the GaAs, the spin can be perpendicular to the momentum $k$ (Fig. 2i).

Interestingly, despite the potential applications in spintronics and valleytronics, the general description of how the little group of $k$-vectors determines the spin-polarization pattern has remained unappreciated. Indeed, the full classification of the spin vector field geometry is an open problem [11]. The common characterization of ST patterns in terms of the CPGS (e.g., the existence or absence of electric polarization [49]) applies only for WPGS $\mathcal{G}(k)$ equal to the CPGS (e.g., the $\Gamma$ point in all lattices, or the Z point in the BZ of GeTe), as shown in Table I. From the examples of ST prototypes in GeTe, Te, and GaAs (Table I and Fig. 2), we can directly anticipate some physical consequences: *i*) The existence of different symmetries of a particular wave vectors $k_0$ in the BZ leads to the possibility of having more than one spin-polarization prototype pattern in the same compound. This suggests that a ST assignment based only on the CPGS can lead to a misclassification of the ST shapes; and *ii*) Compounds without electric dipoles can have *Rashba-like* ST (Fig. 2i), meaning that contrary to what has been traditionally established, the asymmetry of the electric polarization is not a necessary condition for the Rashba ST.

## IV. Derivation of the relation between the point group of the $k$-point and spin-polarization patterns

Examples of the connection between the WPGS in the BZ and macroscopic properties include the notion of "elementary band representation", proposed by Zak [50–52], and its extension [43] including SOC that distinguish compounds having symmetry protected topological phases. Inspired by this idea, we investigate how specific symmetry operations in the WPGS $\mathcal{G}^*(k)$ imposes rules on the $\vec{S}_{nk}$. These symmetry operations contained in WPGS $\mathcal{G}^*(k)$ can be orientation-preserving symmetries termed first kind, such as rotations $R_m$ of an angle $\theta = 2\pi/m$ (with $m$ being an integer), or orientation-reversing symmetries (e.g., reflections $M_i$ and roto-inversions) termed second kind [4]. We use two equivalent approaches to establish the relation between little point groups $\mathcal{G}^*(k)$ and ST shapes for a given wavevector $k_0$, namely: *A.* We derive the linear-in-*k* SOC Hamiltonian $\mathcal{H}(k \to k_0)$ for all possible WPGS in the BZ of non-centrosymmetric compounds. After diagonalizing all Hamiltonian prototypes, the eigenvectors $\psi_n(k)$ are then used to calculate the expectation value $\langle S_{nk_0}(k) \rangle$; and *B.* based on the symmetry transformation properties of pseudo-vectors, we determine how the specific proper and improper symmetry operations determine the direction of the spin-polarization.

## A. *Linear-in-k SOC Hamiltonian for WPGS in the BZ of non-censtrosymmetric compounds*

The symmetry operations contained in a specific WPGS $\mathcal{G}^*(k)$ induce a set of irreducible representations $\gamma_m(d_m)$, which are described in the character table of the point group $\mathcal{G}^*(k)$ [53]. Here, $d_m$ is the dimension of the representation $\gamma_m$, which in turns corresponds to the trace of identity symmetry operation $E$. In Table II, the character table of the point group $C_{3v}$ is represented. Using the symmetry operators in a given basis (the matrix form of the symmetry operators), one can then evaluate under which irreducible representation of $\mathcal{G}^*(k)$, a specific object or property $P$ is transformed. For instance, in table III, the irreducible representations of functions $f_{\gamma_n}(k)$ and Pauli matrices $\sigma_i$ are shown for irreducible representations $\gamma_{1,2} = A_{1,2}$ and $\gamma_3 = E$. In general, bands at the wavevector $k_0$ can also be characterized by the irreducible representations $\gamma_m$ of the point group $\mathcal{G}^*(k_0)$. If a band $n$ transforms under the irreducible representation $\gamma_m$, the dimension $d_m$ correspond to the degeneracy of the given band.

Table II. Character table of the double point group $C_{3v}$ [53]. The headers are the symmetry operations contained in the point group $C_{3v}$, namely: the identity $E$, the threefold rotation symmetry $R_3$, the diagonal mirror plane $M_\sigma$, and the time-reversal symmetry $\tau$. For each irreducible representation ($\gamma = A_1, A_2, E, E_{1/2}, {}^1E_{3/2}, {}^2E_{3/2}$), the characters are shown.

| $C_{3v}$ | $E$ | $2R_3$ | $3M_v$ | $\tau$ |
|---|---|---|---|---|
| $A_1$ | 1 | 1 | 1 | $a$ |
| $A_2$ | 1 | 1 | -1 | $a$ |
| $E$ | 2 | -1 | 0 | $a$ |
| $E_{1/2}$ | 2 | 1 | 0 | $c$ |
| ${}^1E_{3/2}$ | 1 | -1 | $i$ | $b$ |
| ${}^2E_{3/2}$ | 1 | -1 | $-i$ | $b$ |

The Hamiltonian must be invariant, so it transforms under the identity irreducible representations $\gamma_1$– the representation in which all characters are one (e.g., the representation $A_1$ for the point group $C_{3v}$ in Table II). From the tables of the direct product of representations [53], we see that the tensor product $\gamma_m \otimes \gamma_m$ has usually at least one scalar that transform according to identity irreducible representations $\gamma_1$. The Hamiltonian can thus be constructed by considering the sum of products between a function $f_{\gamma_n}(k)$ and a basis matrix $X_{\gamma_m}$ (e.g., Pauli matrices $\sigma_i$ ($i = 0,1,2,3$) are the basis for Hermitical matrices with $n = 2$) that transform under the same irreducible representation $\gamma_m$. For instance, for a two-bands effective Hamiltonian, e.g., one orbital with spin ↑ and ↓, we have,

$$\mathcal{H}(k) = \sum_{m,i} c_m f^i_{\gamma_n}(k) \sigma_i^{\gamma_m} \delta_{nm},$$

where $c_m$ are real coefficients. As an illustrative example of these products, we use the point group $C_{3v}$. For example, $\sigma_z$ transforms under the irreducible representation $\gamma_2 = A_2$ (Table III) and there are no functions $f_{A_2}(k)$ that transform under the representation $\gamma_2$, so there are no terms obtained from the product $\gamma_2 \otimes \gamma_2$ in the Hamiltonian. Considering all products containing Pauli matrices that are even under the time-reversal symmetry, the resulting Hamiltonian for $\mathcal{G}^*(k_0) = C_{3v}$ is given by:

$$\mathcal{H}(k \to k_0) = c_1 \mathbb{1} k^2 + c_3 (k_x \sigma_y - k_y \sigma_x),  \quad (1)$$

where $k^2 = k_x^2 + k_y^2 + k_z^2$. The physical interpretation of the Hamiltonian allows to directly determine the coefficients $c_m$. The first term of the Hamiltonian is obtained from $\gamma_1 \otimes \gamma_1$ (with $\gamma_1 = A_1$) and gives the kinetic energy, meaning that $c_1 = -\hbar/2m^*$. Similarly, the second term in Eq. 1 (obtained from $\gamma_3 \otimes \gamma_3$ with $\gamma_3 = E$) corresponds to the linear-in-$k$ Rashba SOC Hamiltonian $\mathcal{H}_R$ with $c_3 = \lambda_R$. This simple analysis not only allows to determine that the Rashba Hamiltonian $\mathcal{H}_R$ is symmetrically allowed in the WPGS $\mathcal{G}^*(k_0) = C_{3v}$, but it also indicates that the Weyl and Dresselhaus Hamiltonians are symmetrically forbidden. A detailed description of this method can be found in Ref [22]. This method of invariants has been used to study the linear-in-$k$ Rashba SOC Hamiltonians allowed by the symmetry operations of polar point groups, as well as the high-order contributions to the Rashba-Bychkov effect [54,55]. Here, we extend this approach to determine the effective SOC Hamiltonian for all possible wavevector point groups $\mathcal{G}^*(k_0)$ that are non-centrosymmetric (NCS).

Table III. Cartesian tensors $\sigma_i$ and $s$, $p$, $d$, and $f$ functions ($l = 0, 1, 2, 3$) [53].

| $C_{3v}$ | 0 | 1 | 2 | 3 |
|---|---|---|---|---|
| $A_1$ | $\mathbb{1}_{2\times 2}$ | $k_z$ | $k_x^2 + k_y^2, k_z^2$ | $k_x(3k_x^2 - k_y^2), k_z(k_x^2 + k_y^2), k_z^2$ |
| $A_2$ | | $\sigma_z$ | | $y(3k_x^2 - k_y^2)$ |
| $E$ | | $(k_x, k_y)(\sigma_y, -\sigma_x)$ | $(k_x k_y, k_x^2 - k_y^2), (k_z k_x, k_y k_z)$ | $\{k_x(k_x^2 + k_y^2), k_y(k_x^2 + k_y^2)\}$, $k_z^2(k_x, k_y)$, $\{k_x k_y k_z, k_z(k_x^2 - k_y^2)\}$ |

Figure 3 summarizes all SOC Hamiltonians that are symmetrically allowed by the specific WPGSs $\mathcal{G}^*(k_0)$, which according to the polarity and chirality are classified in four categories, namely: polar chiral ($C_1, C_2, C_3, C_4$, and $C_6$), polar non-chiral ($C_s, C_{2v}, C_{3v}, C_{4v}$, and $C_{6v}$), non-polar chiral ($D_2, D_3, D_4, D_6, T$, and $O$), and non-polar non-chiral ($D_{2d}, S_4, C_{3h}, D_{3h}$, and $T_d$). These four categories are based on the existence of specific proper (rotations) and improper symmetries (reflections and inversion) in the wavevector point group symmetry. Specifically, polar and non-polar PGs have a single and more than one rotation axis, respectively. On the other hand, chiral PGs have only proper symmetries, while non-chiral have both proper and improper symmetries. Thus, the WPGS classification according to polarity and chirality groups the kind of symmetry operations contained in the WPGS, which has implications in the symmetry enforced ST shapes. Specifically, we identify two extreme behaviors for the ST, i.e., $\vec{S}_{nk} \perp \vec{k}$ (*tangential* ST) and $\vec{S}_{nk} \parallel \vec{k}$ (*radial* ST), resulting from the diagonalization of the Hamiltonians $\mathcal{H}_R$ and $\mathcal{H}_W$, respectively. The other possible STs are combinations of these two extreme behaviors as in the *radial-tangential* ST associated to the Hamiltonian $\mathcal{H}_D$. As summarized in Figure 3, there is a trend in the symmetrically allowed Hamiltonians for the four WPGS categories: (a) polar chiral WPGS $C_3, C_4$, and $C_6$ symmetrically allow the Hamiltonians $\mathcal{H}_R$ and $\mathcal{H}_W$, while in $C_1$ and $C_2$ have no constrains; (b) polar non-chiral WPGS $C_{3v}, C_{4v}$, and $C_{6v}$ only allow the Rashba Hamiltonian, while polar non-chiral point groups $C_s$ and $C_{2v}$ lead to the SOC terms $\mathcal{H}_R$ and $\mathcal{H}_D$; (c) non-polar chiral WPGS $D_3, D_4, D_6, T$, and $O$ leads to the SOC term $\mathcal{H}_w$, but the non-polar chiral WPGS $D_2$ leads to both $\mathcal{H}_w$ and $\mathcal{H}_D$; and (d) all non-polar non-chiral WPGS $D_{2d}, S_4, C_{3h}, D_{3h}$, and $T_d$ symmetrically impose the Hamiltonian $\mathcal{H}_D$. Additional symmetry constraints can be imposed by high-order-in-$k$ SOC terms, which usually results from functions $f_{\gamma_m}(k)$ related to $d$ and $f$ functions (e.g., functions in columns with $l > 1$ in Table III). Since we focus here only on non-magnetic compounds, i.e., compounds preserving the time-reversal symmetry $\mathcal{T}$, the ST in anti-ferromagnetic compounds [56] as well as the Zeeman-type effect at $k_0$ breaking $\mathcal{T}$ [8,9] are not included. Thus, all ST prototypes above described are assumed to intrinsically satisfy the condition $\vec{S}_{n,-k} = \mathcal{T}\vec{S}_{n,k} = -\vec{S}_{n,k}$ (i.e., the pseudo-vector $\vec{S}_{nk}$ at the inverted $k$-vector is also inverted). In general, the SOC Hamiltonian terms allowed at $k_0$ [55] intrinsically satisfy the symmetry constraints to the ST, as shown in the next phenomenological discussion.

| C/NC<br>P/NP | Chiral (C) WPGSs | | Non-Chiral (NC) WPGSs | |
|---|---|---|---|---|
| **Polar (P) WPGSs** | $C_1$ | No constrain | $C_s$ | Rashba and Dresselhaus |
| | $C_2$ | No constrain | $C_{2v}$ | Rashba and Dresselhaus |
| | $C_3$ | Rashba and Weyl | $C_{3v}$ | Rashba |
| | $C_4$ | Rashba and Weyl | $C_{4v}$ | Rashba |
| | $C_6$ | Rashba and Weyl | $C_{6v}$ | Rashba |
| **Non-polar (NP) WPGSs** | $D_2$ | Dresselhaus and Weyl | $S_4$ | Dresselhaus |
| | $D_3$ | Weyl | $C_{3h}$ | Dresselhaus |
| | $D_4$ | Weyl | $D_{2d}$ | Dresselhaus |
| | $D_6$ | Weyl | $D_{3h}$ | Dresselhaus |
| | $T$ | Weyl | $T_d$ | Dresselhaus |
| | $O$ | Weyl | | |

Figure 3: Classification of the WPGS and the respective linear-in-$k$ Hamiltonian that is symmetrically allowed by the polarity and chirality of the WPGS.

**B.** *Analysis of how the specific proper and improper symmetry operations determine the direction of the spin-polarization*

Using the symmetry constrains imposed by the symmetry operations contained in the WPGS $\mathcal{G}^*(k)$, we identify the allowed spin-polarization patterns in the BZ of NCS compounds. This phenomenological analysis is based on the pseudo-vector properties of $\langle S_{nk_0}(k) \rangle$.

If a vector property $\vec{p}$, such as the ferroelectric polarization, is perpendicular to a rotation axis, only $R_m$ with $m = 1$ preserves the crystal unchanged (i.e., rotation of $\theta = 2\pi/m$, which by definition is the identity operation $E$). Equivalently, if $R_m$ with $m \neq 1$ is a symmetry operation of the crystal, the physical property $\vec{p}$ must be parallel to the rotation axis. The spin-polarization $\vec{S}_{nk}$ is a pseudo-vector that locally must be parallel to the rotation symmetry operations contained in the WPGS $\mathcal{G}^*(k)$. On the other hand, if a reflection plane $M_i$ is a symmetry operation of $\mathcal{G}^*(k)$, a pseudo-vector (vector) property $\vec{p}$ must be perpendicular (parallel) to the mirror plane. Thus, $\vec{S}_{nk}$ must be perpendicular to the rotation symmetry operations contained in the WPGS $\mathcal{G}^*(k)$. The SOC Hamiltonian terms symmetrically allowed at $k_0$ [55] intrinsically satisfy above-noted constraints imposed by rotations $R_m$ and mirror symmetries $M_i$ in the ST (i.e., $\vec{S}_{nk} \perp M_i$ and $\vec{S}_{nk} \parallel R_i$). Interestingly, the four WPGS categories defined by the polarity and chirality lead to group of rules (i.e., symmetry constrains) for the spin-polarization vector $\vec{S}_{nk_0}$, as described in the four respective quadrants (a)-(d) in Figure 4. To illustrate the relations between categories of little point group symmetry $\mathcal{G}^*(k_0)$ of a particular wavevector $k_0$ and the spin-polarization pattern around it, we explain below how simultaneous symmetry restrictions such as $\vec{S}_{nk} \perp M_i$ and $\vec{S}_{nk} \parallel R_i$ can determine the ST described by the specific SOC terms $\mathcal{H}_R$, $\mathcal{H}_w$, and $\mathcal{H}_D$:

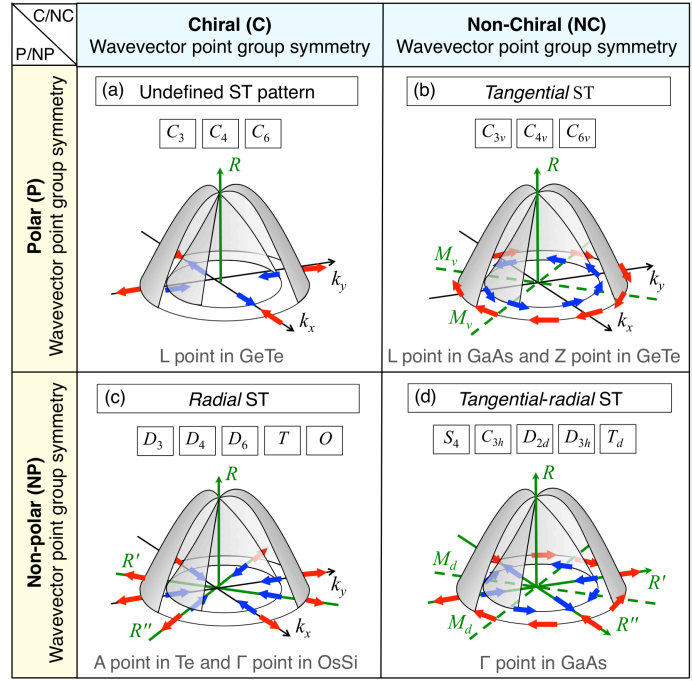

Figure 4. Classification of the WPGS and STs according to the polarity and chirality of the WPGS. Blue and red arrows represent the ST of the inner and outer energy contour of spin split bands. Green solid and green dashed lines stand for the rotation axis and mirror plane projections in a plane perpendicular to the rotation axis, respectively. In each panel, examples of ST at different high-symmetry wavevectors are presented. In the specific case of the WPGSs $C_1$ and $C_2$ (polar chiral WPGS – panel (a)), there is no defined ST pattern, as shown in Fig. 3. For $C_s$ (only one mirror plane) and $C_{2v}$ (no mirror planes), which are polar non-chiral WPGS (panel (b)), a combination of ST prototypes with near zero out-plane ST is expected, rather than a perfect *tangential* ST.

*(a) k points having polar chiral WPGSs can be characterized by the absence of limiting spin textures:* In the polar chiral WPGS (Fig. 3a), there is only one rotation axis $R_m$ (a polar axis). As a result there are no constrains on the in-plane directions of the polarization $\vec{S}_{nk}$ (Fig. 4a). We refer to this condition as the *absence of pure limiting ST behaviors* [either $\vec{S}_{nk} \perp \vec{k}$ or $\vec{S}_{nk} \parallel \vec{k}$]. Group symmetry analysis indicate that for $C_3, C_4$, and $C_6$, the linear-in-$k$ SOC Hamiltonian at $k_0$ can be written as a superposition of the Rashba $\mathcal{H}_R$ and Weyl $\mathcal{H}_w$ SOC terms [55]. The spin-polarization pattern observed in WPGS $\mathcal{G}^*(k_0) = C_m$ (with $m$= 3, 4, and 6) depend on the strength of Rashba and Weyl SOC terms $\lambda_R$ and $\lambda_w$, respectively (all combinations between $\lambda_R$ and $\lambda_w$ are symmetry allowed). Rashba and Weyl SOC intrinsically imposes that at the rotated momentum vector $\vec{k}$ (i.e., $\hat{R}_m \vec{k}$), the spin-polarization $\vec{S}_{nk}$ near $k_0$ is also rotated, i.e., $\hat{R}_m \vec{S}_{nk}$ (with $\hat{R}_m$ being the rotation symmetry operator), as required by the existence of the polar axis $R_m$. In figure 4a, the ST expected in *k*-points with point groups having rotation symmetry $R_4$ (i.e, $C_4$) is illustrated. Due to the *absence of limiting behaviors* (i.e., pure Rashba or Weyl STs) and possible arbitrary $\lambda_R$ and $\lambda_w$, in polar chiral WPGS, there are no defined patterns for the in-plane spin-polarization.

*(b) k points having polar non-chiral WPGS can be characterized by pure Rashba spin texture:* In $k_0$ with polar non-chiral WPGS (Fig. 3b), there is one polar rotation axis $R_n$ and vertical mirror planes ($M_v$), i.e., planes containing the polar axis. For instance, the $\mathcal{G}^*(k_0) = C_{3v}$ is formed by threefold rotation symmetries $R_3$ and three vertical mirror planes $\sigma_v$ (i.e., the planes containing the rotation axis $R_3$). Thus, $\vec{S}_{nk}$ is required to be perpendicular to the planes $M_v$ (as represented at the green dashed lines in Fig. 4b). Like in polar chiral WPGS, the spin-polarization pattern is also required to satisfy the rotation symmetry at all *k*-points around the polar axis. Thus, the coexistence of polar rotation symmetries and vertical mirror planes implies that $\vec{S}_{nk}$ is perpendicular to the momentum $\vec{k}$ ($\vec{S}_{nk} \perp \vec{k}$), which is referred to as spin-momentum locking effect. This effect characterizing the "Rashba- ST" is enforced by symmetry rather than a consequence of the magnitude of the SOC or electric dipoles. In summary, non-Sohncke polar PGs are required to have $\vec{S}_{nk} \perp \vec{k}$ – the pure Rashba ST. The Hamiltonian describing this extreme behavior is then given by the "pure" Rashba SOC term $\mathcal{H}_R$.

In actual compounds, high symmetry *k*-points having $\mathcal{G}^*(k_0) = C_{3v}, C_{4v}$, and $C_{6v}$ are expected to have this limiting behavior. Examples include $k_0 = $ L in GaAs (*F$\bar{4}$3m*) and $k_0 = $ Z in GeTe (*R3m*), as shown in Fig. 3c and 3e, respectively. Although first predicted for surfaces with perpendicular external field, the Rashba ST has recently been generalized for bulk compounds, which has motivated the search of this compounds with large spin splitting [7].

*(c) k points having non-polar chiral WPGS can be characterized by pure Weyl spin texture*: For $\mathcal{G}^*(k_0) = D_2, D_3, D_4, D_6, T$, or $O$ (Fig. 3c), there is a single polar rotation axis $R_m$ and at least one additional rotation axis ($R'_m$ and $R''_m$) lying in the plane perpendicular to the polar rotation axis $R_m$. Here $\vec{S}_{nk}$ is then required to be parallel to the rotation axis $R'_m$ and $R''_m$ (as represented at the green solid lines in Fig. 3c). Additionally, at the rotated momentum $\vec{k}$ (corresponding to $\hat{R}_m\vec{k}$ or $\hat{R}'_m\vec{k}$), the pseudo-vector $\vec{S}_{nk}$ is also rotated ($\hat{R}_m\vec{S}_{nk}$ or $\hat{R}'_m\vec{S}_{nk}$). The existence of rotation symmetries perpendicular to the polar axis $R_m$ implies that $\vec{S}_{nk}$ is parallel to the momentum $\vec{k}$ (as in $\vec{S}_{nk} \parallel \vec{k}$), as shown in Fig. 3c. This *radial* ST (Fig. 1c) is locally given by SOC Weyl term $\mathcal{H}_w$. The "*Weyl* ST" is usually associated to topological Weyl semimetal [36,37] and Kramers-Weyl fermions in chiral compounds [10–12].

In actual compounds, high symmetry *k*-point having non-polar chiral WPGS include the $\Gamma$ and A point in chiral Te bulk [10–12], as well as the $\Gamma$ point of the insulators OsSi (*P2$_1$3*) and SeF$_4$ (*P2$_1$2$_1$2$_1$*). As predicted by our description, these k-points indeed have the *Weyl ST* (Fig. 2h). In general, all non-polar chiral compounds can have the Weyl ST prototype around the $\Gamma$ point (or other k-point whose WPGS is equal to the CPGS).

*(d) k points having non-polar non-chiral WPGS will show a mixture of pure spin textures:* The ST around *k*-point having WPGS $\mathcal{G}^*(k_0) = D_{2d}, S_4, C_{3h}, D_{3h}$, or $T_d$ can be seen as a combination of the Weyl and Rashba STs, as shown in Fig. 4d. The reason is that these PGs contain one polar axis ($R_m$), additional rotation axes ($R'_m$ and $R''_m$) perpendicular to $R_m$ as required by Weyl ST, and mirror planes as required by Rashba ST. These reflection planes can be perpendicular to the polar rotation axis (horizontal mirror planes $M_h$) or can bisect the angle between a pair of rotational axis (diagonal mirror planes $M_d$). Thus, $\vec{S}_{nk}$ is required to be parallel to the in-plane rotation axis $R'_m$ and $R''_m$ and also perpendicular to mirror planes $M_d$ and $M_h$, leading to a combination of the extreme behaviors $\vec{S}_{nk} \perp \vec{k}$ and $\vec{S}_{nk} \parallel \vec{k}$ (e.g., the $\Gamma$ point of GaAs, which is described by the SOC term $\mathcal{H}_D$).

## C. *Some consequences of the relation between the k-point point group and spin texture prototypes*

The above-noted ST classification and the existence of symmetries of particular wave vectors $k_0$ in the BZ leads to the possibility of having more than one spin-polarization prototype pattern in the same compound. Some additional consequences include: (i) Compounds without electric dipoles can have the *tangential* ST (Rashba ST), meaning that contrary to what has been traditionally established, the asymmetry of the electric polarization is not a necessary condition for the Rashba effect (Fig. 2c); (ii) compound with electric dipoles can have the Dresselhaus SOC term $\mathcal{H}_D$ at polar chiral WPGS (Fig. 2f); (iii) Spin split bands can have vanishing ST. The *tangential-radial* ST is not the only way to combine pure Rashba $\vec{S}_{nk} \perp \vec{k}$ and Weyl $\vec{S}_{nk} \parallel \vec{k}$ STs. For instance, when a rotation axis $R_m$ is contained in a diagonal $M_d$ or horizontal $M_h$ mirror plane, the spin-polarization is simultaneously imposed to be parallel and perpendicular to the rotation axis. This contradiction implies that the pseudo-vector $\vec{S}_{nk}$ must vanish (even in spin split bands), as in two-dimensional SnTe thin film [57]; and (iv) another interesting consequence is the symmetry-enforced *radial* ST ($\vec{S}_{nk} \parallel \vec{k}$) in bulk compounds. This ST is believed to be a characteristic only linked to symmetry protected topological phases [36,37]. We find that polar chiral compounds can possess a radial ST $\vec{S}_{nk} \parallel \vec{k}$ at the $\Gamma$ point, even without symmetry protected topological phase. For instance, Figure 5 shows DFT ST for the valence band maximum at the $\Gamma$ point of OsSi ($P2_13$). Our finding explain the experimental observation via spin-angle-resolved photoemission spectroscopy of the *radial* ST in Bulk Te [10–12].

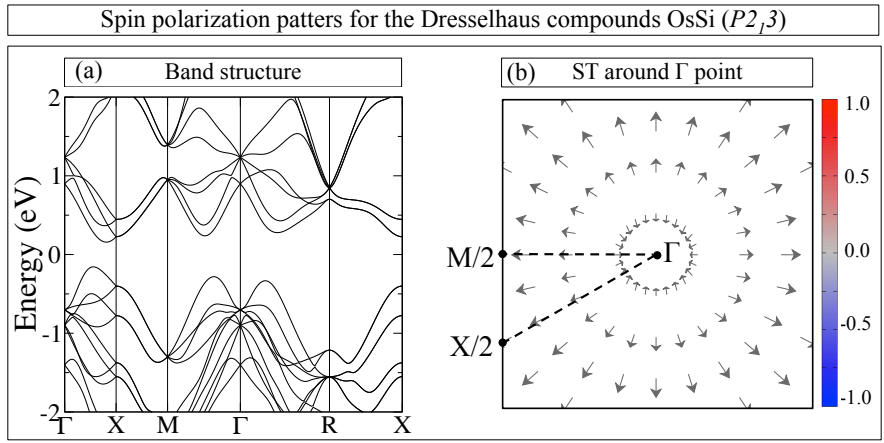

Figure 5. Relativistic density functional band structure (a) for the "Dresselhaus compound" OsSi ($P2_13$) and (b) Spin texture (ST) for the CBM at the $\Gamma$ point showing typical Weyl-like texture

Figures 3 and 4 provide a complete classification of the ST in terms of the symmetry conditions intrinsically imposed by WPGSs without symmetry inversion. Although we can directly identify some consequences from the ST classification, the relation between WPGS and ST do not establish all STs that can be find in single compound with a given CPGS. Indeed, one may erroneously believe that any compound can have any ST shape. In the next section, we thus complete this description by relating the CPGS to a set of symmetrically imposed STs.

## V. Description of the design principles imposed by the wavevector point group symmetries on the spin-texture of a single compound

The developed classification for ST shapes for WPGSs without inversion symmetry ratifies that the ST for wavevectors $k_0$ at which WPGS=CPGS can be predicted by the CPGS. However, this classification does not disclose all ST shapes that can be observed near high-symmetry $k$-points in the BZ of a given NCS crystalline compound with a specific CPGS $\mathcal{G}$. This is a fundamental problem for potential spintronic and valleytronic applications. In order to answer this question, we examine the subgroups for all CPGS $\mathcal{G}$ without inversion symmetry. As very well established, for a given point group $\mathcal{G}$, there is a limited set of subgroups, which are imposed to have lower symmetry than $\mathcal{G}$, e.g., the subgroups of the point group $T_d$ are $C_1$, $C_2$, $C_s$, $C_{2v}$, $C_{3v}$, $D_2$, $T$, $S_4$, and $D_{2d}$. Since WPGSs are subgroups of the CPGS (including also the case WPGS=CPGS), the CPGS imposes a limited set of ST shapes in the BZ. The matrix illustrated in Fig. 6 summarizes the symmetrically allowed STs for each CPGS. The columns and lines stand for the CPGS and WPGS classified according to the polarity and chirality, respectively. The matrix component corresponding to the intersection of a CPGS with a WPGS are yellow (gray) when the WPGS is (is not) a subgroup of the CPGS. In other words, the yellow components indicate the WPGS that can exist in the BZ of a compound having any of the considered CPGS. For instance, since the subgroups have lower symmetry than the point group, there is no high-symmetry $k$-point with WPGS that have larger symmetry than the CPGS and hence, the lower triangle of the square matrix in Fig. 6 is completely gray. When the WPGS lead to a ST with a defined shape, we include letters indicating the symmetrically allowed linear-in-$k$ SOC Hamiltonians: Rashba (R), Dresselhaus (D), Weyl (W), Rashba and Weyl (RW), Rashba and Dresselhaus (RD), and Weyl and Dresselhaus (WD).

As illustrated in Fig. 6, the *design principles* (DP) for multiple ST shapes in a single compound can be summarized as follows:
  a. Polar chiral CPGS could only have high-symmetry $k$-points with polar chiral WPGS (first column in Fig. 6). Thus, in compounds with these CPGS, the ST has no *pure limiting ST behaviors* [either $\vec{S}_{nk} \perp \vec{k}$ or $\vec{S}_{nk} \parallel \vec{k}$], resulting from the superposition of the SOC terms $\mathcal{H}_R$ and $\mathcal{H}_W$.
  b. Polar non-chiral CPGS could have high-symmetry $k$-points with polar WPGS that are chiral and non-chiral (second column in Fig. 6). The *tangential* ST, imposed by Rashba Hamiltonian $\mathcal{H}_R$, is thus the only limiting behavior that can be observed in polar non-chiral compounds. In these compounds with CPGSs $C_s$ and $C_{2v}$, the ST is a combination of the patterns arising from Rashba and Dresselhaus SOC terms ($\mathcal{H}_R + \mathcal{H}_D$). Besides this ST, in polar non-chiral compounds with CPGSs $C_{3v}$, $C_{4v}$, and $C_{6v}$, it is also possible to have a ST arising from the simultaneous presence of Rashba and Weyl SOC terms ($\mathcal{H}_R + \mathcal{H}_W$).

c. Non-polar chiral CPGS can have high-symmetry $k$-points with chiral WPGS that are polar and non-polar (third column in Fig. 6). The *radial* ST, imposed by the Weyl Hamiltonian $\mathcal{H}_W$, is the only limiting behavior in non-polar chiral compounds. Additionally, compounds with CPGSs $D_3$, $D_4$, $D_6$, $T$, and $O$ can also have ST arising from the simultaneous presence of Rashba and Weyl SOC terms ($\mathcal{H}_R + \mathcal{H}_W$), as well as Dresselhaus and Weyl SOC terms ($\mathcal{H}_D + \mathcal{H}_W$).

d. Non-polar non-chiral CPGS can have WPGS with all possible combinations of polarity and chirality (four column in Fig. 6). The *radial-tangential* ST can be found in all non-polar chiral compounds. Additionally, both limiting behaviors for *tangential* and *radial* STs could be observed in non-polar chiral compounds.

Figure 6. Classification of CPGS (columns) and WPGS (lines) according to the polarity and chirality. The intersection between CPGS and WPGS are discriminated by yellow and gray colors. Yellow (gray) means that the specific WPGS is (is not) a subgroup of the CPGS $\mathcal{G}$ and hence, the WPGS can (cannot) be in the BZ of a compound having crystal point group $\mathcal{G}$. For each matrix element, we specify the linear-in-$k$ SOC Hamiltonian symmetrically allowed by the WPGS: Rashba (R), Dresselhaus (D), Weyl (W), Rashba and Weyl (RW), Rashba and Dresselhaus (RD), and Weyl and Dresselhaus (WD).

According to this description, only polar non-chiral, non-polar chiral and non-polar non-chiral CGPS can have multiple ST in the BZ. In order to illustrate these design principles (a-d), we study the DFT calculated valley-dependent ST in GaAs as represented in Fig. 7. These compounds have $T_d$ CPGS, and hence, the PG of high-symmetry $k$-points in the BZ of GaAs correspond to the subgroups of $T_d$ (i.e., $C_s, C_2, C_3, D_2, C_{2v}, C_{3v}, D_{2d}, S_4$, and $T$). However, there is no unequivocal correspondence between the number of high-symmetry *k*-points and the number of subgroups of the lattice. For instance, high symmetry *k*-points Γ, X, L, W, K, and U have PGs $T_d$, $D_{2d}$, $C_{3v}$, $S_4$, $C_s$, and $C_s$ (see Fig. 2g), respectively, meaning that there is no *k*-point in the BZ of GaAs with PG symmetry *T*, as illustrated by hierarchical decomposition of the subgroups of the PG $T_d$ represented in Fig. 7. The little PGs of *k*-points impose specific STs, represented only for high symmetry *k*-points Γ, X, and L (Fig. 7): the L point is required to have spin-polarization perpendicular to the momentum *k* ($\vec{S}_{nk} \perp \vec{k}$), and the *Γ* point can have a mixing of the extreme behaviors $\vec{S}_{nk} \perp \vec{k}$ and $\vec{S}_{nk} \parallel \vec{k}$. In other words, while the *Γ* point in GaAs has the *tangential-radial* ST, as expected, the L point has *radial*-like ST (Fig. 2h-i). Similarly, the high symmetry points X, W, and K have ST (not shown in Fig. 7) similar to the one expected at the *Γ* point.

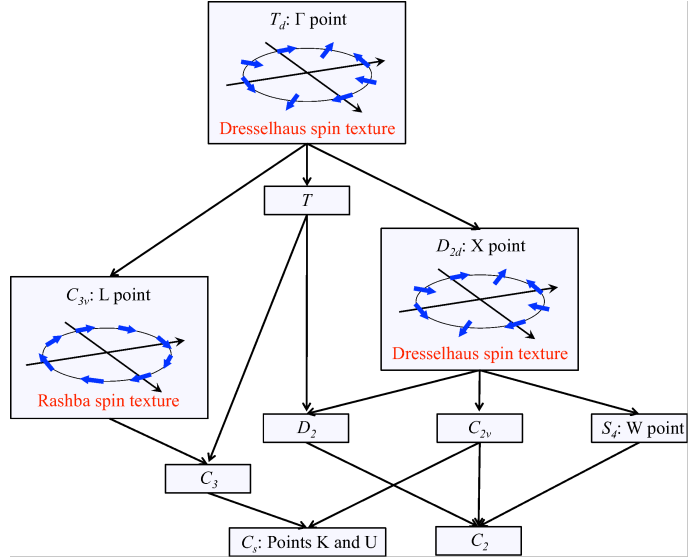

Figure 7. The Bärnighausen tree provides a hierarchical decomposition of PG $T_d$ into its subgroups (i.e., $C_s, C_2, C_3, D_2, C_{2v}, C_{3v}, D_{2d}, S_4$, and $T$), e.g., PG $T_d$ can be decomposed in the lower symmetry PGs $C_{3v}, D_{2d}$, and $T$. Subgroups $T$ can in turn be decomposed into PGs $C_3$ and $D_2$. For each PG, the high-symmetry *k*-points are specified, i.e., Γ, X, L, W, K, and U for the PGs $T_d$, $D_{2d}$, $C_{3v}$, $S_4$, $C_s$, and $C_s$, respectively. The spin texture expected by the subgroup is represented for Γ, X, and L points.

## VI. DFT illustrations of the journey through the Brillouin Zone and high-throughput calculations

The DPs establishing the relationship between each non-magnetic NCS CPGS and the possible ST shapes (Fig. 6) gives the possibility of rationally selecting compounds that have different spin textures in the BZ. In this section, we apply the previously described theory for the design of compounds that can potentially be used in spintronic devices. Besides the symmetry conditions established in Fig. 6 (i.e., enabling design principles (EDP)), we also consider DPs for the optimization of the target functionalities, e.g., multiple ST shapes, its position in the BZ and energy. Optimizing DPs (ODP) depend on the specific application. Here, we focus on spin-valleytronics, a rapid growing area based on the use of the spin-polarization patterns that are controllable by the degree of freedom of the *k*-valleys in the BZ. Figure 8 summarizes the EDPs for different ST shapes in the same compound as well as ODPs for the control of these STs shapes for spin-valleytronics. Specifically, the starting point is a list of materials that are non-magnetic and gapped. The EDPs for compounds having multiple ST shapes in the BZ are (i) EDP1: NCS CPGS, and (ii) EDP2: CPGS should also be polar non-chiral, non-polar chiral or non-polar non-chiral. Finally, as we discuss below, the ODP includes ODP1: linear-in-$k$ SSs larger than 1 meV at band edges and $k$-valleys with different WPGSs, and ODP2: enough small energy difference between states at different valleys. After selecting compounds using the DPs as filters (Tables V and VI), we focus on the illustration of some specific prototypes.

### A. Inverse design of compounds with multiple spin texture shapes that can potentially be controlled by the valley degree of freedom

In spin-valleytronics, one wants to have an association between valleys and spin-polarization as established in previous sections based on the WPGS. The basic idea is that if two high symmetry $k_1$ and $k_2$ in the BZ have different WPGS, the ST shapes can also be different around these wavevectors. In experiments, $k_1$ and $k_2$ can independently be accessed in order to select different STs. Additionally, since the electronic transport and spin-currents are governed by the electronic states near the band edges, the spin-currents also depend on the WPGS of the $k$-point at which the band edges take place. The controllable valley energy requires a sufficiently small energy difference between states at different valleys. The use of this ODP as a filter requires the evaluation of the DFT SOC band structure for a relatively large set of compounds. For this reason, in order to reduce the computation cost, before applying the EDPs, we delimit the list of compounds by selecting non-magnetic gapped materials based on previous DFT calculations without SOC, as described below. Below, we describe the three steps of the materials selection process, namely: materials filtering based on the previous DFT calculations, materials selection based on the symmetry conditions (i.e., enabling DPs), and materials optimization (i.e., optimizing DPs).

#### 1. Find the subset of materials that are non-magnetic gapped compounds

In order to reduce the computational cost of high-throughput density functional calculations, we delimit the studied compounds according to their atomic features, i.e., the number of atoms in the unit cell and the orbital type. Our starting point is thus a list from the aflow-ICSD database containing **20,831** unique compounds with less than 20 atoms per unit cell [29] and restricted to atoms having only *s*, *p* and *d* orbitals. In the aflow-ICSD database, there were initially 58,276 entries (32,115 removing duplicated entries). Since we focus on gapped compounds preserving the time-reversal symmetry, we restrict the materials selection to non-magnetic compounds with non-zero bandgap. The screening of non-magnetic insulators has the bias of the DFT calculations performed in the aflow-ICSD database, where the charge density is usually initialized with a ferromagnetic configuration. This could make anti-ferromagnetic compounds to be reported as ferromagnetic. In the aflow-ICSD database, we use the *spin_cell* feature, which correspond to the total magnetic moment per unit cell, to filter nonmagnetic compounds. This materials screening divides the initial database in two groups: 6,993 magnetic materials and 13,838 non-magnetic compounds. On the other hand, in the aflow-ICSD, non-spin-polarized calculations classify compounds as direct gap insulators, indirect gap insulators, metals, and half metals. Based on this classification, the 13,838 non-magnetic compounds are then divided into 7,483 non-gapped and 6,355 gapped compounds (i.e., band gap larger that 1 meV), as represented in **line 1** of Fig. 8. These 6355 non-magnetic insulator were previously obtained by us [7,20].

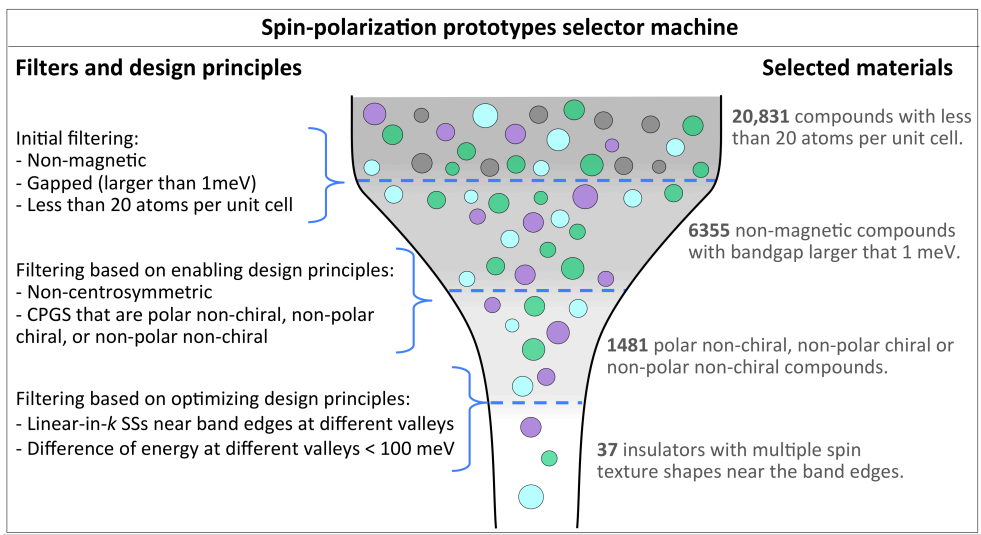

Figure 8. Data-mining approach to select materials with multiple spin texture shapes near the band edges. The different filters and design principles are indicated in the left, and the number of selected materials for each filter are shown in the right side of the sketch.

2. *Find the subset of non-magnetic gapped compounds that can have different ST in the BZ*

We use the crystal point group of the compounds to filter materials satisfying EDP1 and EDP2. Table IV presents the space group index for each CPGS for different polarity, chirality, and Bravais lattices. We notice that there are Bravais lattices with symmetry forbidden CPGS categories. For instance, in orthorhombic and cubic (triclinic and cubic) lattices, there are no polar chiral (non-chiral) CPGS, and in triclinic and monoclinic (triclinic, monoclinic, orthorhombic, and rhombohedral) lattices, there are no non-polar chiral (non-chiral) CPGS. All these point groups are necessarily NCS (EDP1 in Fig. 8). We thus filter from the list of non-magnetic insulators those compounds having NCS CPGS. We find **1709** compounds with NCS CPGSs and other list of 4645 compounds with CS crystal point groups, as represented in **line 2** of Fig. 8. The 1709 compounds with NCS CPGS are divided into **228** polar chiral, **723** polar non-chiral, **253** non-polar chiral, and **505** non-polar non-chiral compounds. In previous papers, we study the spin splitting and ferroelectric properties of the 951 polar (chiral and non-chiral) compounds [7,8,20]. In Table IV, we specify the abundance of compounds for each NCS CPGS. Curiously, there are no non-magnetic insulators having the NCS non-polar chiral CPGS $O$. Besides the CPGS $C_{2v}$ with 320 compounds, the point groups $D_{2d}$, $T_d$, and $C_{3v}$ are the most abundant NCS CPGS with 194, 183, and 176 compounds, respectively (Table IV). Selecting compounds that can have multiple ST in the BZ (i.e., those with polar non-chiral, non-polar chiral or non-polar non-chiral CPGS), we obtain 1481 compounds (EDP2 in Fig. 8).

Table IV. Space group indexes for 3D Bravais lattices classified according to the inversion symmetry as centrosymmetric, NCS non-polar, and NCS polar.

| | | Space groups indexes for the 3D Bravais lattices | |
|---|---|---|---|
| | | **Chiral** | **Non-chiral** |
| **Polar** | Triclinic | $C_1$: 1 (48) | -- |
| | Monoclinic | $C_2$: 3-5 (131) | $C_s$: 6-9 (112) |
| | Orthorhombic | -- | $C_{2v}$: 25-46 (320) |
| | Tetragonal | $C_4$: 75-80 (7) | $C_{4v}$: 99-110 (31) |
| | Rhombohedral | $C_3$: 143-146 (30) | $C_{3v}$: 156-161 (176) |
| | Hexagonal | $C_6$: 168-173 (12) | $C_{6v}$: 183-186 (84) |
| | Cubic | -- | -- |
| **Non-polar** | Triclinic | -- | -- |
| | Monoclinic | -- | -- |
| | Orthorhombic | $D_2$: 16-24 (88) | -- |
| | Tetragonal | $D_4$: 89-98 (22) | $S_4$: 81-82 (68) |
| | | | $D_{2d}$: 111-122 (194) |
| | Rhombohedral | $D_3$: 149-155 (87) | -- |
| | Hexagonal | $D_6$: 177-182 (18) | $C_{3h}$: 174 (4) |
| | | | $D_{3h}$: 187-190 (56) |
| | Cubic | $T$: 195-199 (38) | $T_d$: 215-220 (183) |
| | | $O$: 207-214 (0) | |

Although polar compounds usually have an intrinsic electric polarization, polar CPGS is as necessary but not sufficient condition for electric polarization. The cancelation of dipoles in polar compounds can be geometrically determined for each atomic site by considering vectors along the atomic bonds. Specifically, the electron transfer given by the atomic bonding of two different elements creates a microscopic dipole whose direction is opposite to electron transfer direction. For a given atomic site, if all neighbor atoms were locally distributed in such a way that the dipole vectors generated by each bonding cancel each other, then the local atomic site dipole would be zero, e.g., non-polar atomic sites. Local dipoles can add up to zero for first or second neighbors or be near zero dipole for more distant atomic neighbors (e.g., atomic distances larger than the sum of Van der Waals radius for two given elements), which can also be verified geometrically. Thus, polar compounds with non-polar atomic sites cannot have a total non-zero electric dipole, i.e., non-zero local dipoles can only be found in compounds with polar atomic sites. This gives an intuitive way to verify the local cancelation of dipoles using only the atomic positions and lattice vectors. From the 1481 NCS non-magnetic insulator, 1018 compounds have at least one polar site. We identify **867** compounds with non-zero dipole and **151** compounds in which local dipoles cancel each other. The proposed approach based on the geometrical information can result in false positives non-zero dipole, since the electron dipole also depends on the specific chemical species, which requires more exhaustive first principles calculations. Therefore, we retain this list of 1481 non-magnetic NCS insulators for our next step.

3. *Find the subset of non-magnetic NCS gapped compounds with ST that can potentially be controlled by the valley*

Since controllable valley energy requires a sufficiently small energy difference between states at different valleys $\Delta_{k_1 k_2}$, we restrict the materials selection based on this ODP (ODP2 in Fig. 8). In order to evaluate $\Delta_{k_1 k_2}$, we perform high-throughput DFT band structure calculations for the previously selected 1481 non-magnetic NCS insulators with CPGS allowing multiple ST shapes in the BZ. The DFT calculations with SOC are performed using Perdew-Burke-Ernzenhof generalized gradient approximation (PBE) [58] as the exchange-correlation functional and the Coulomb self-repulsion on-site term U for transition metals [59] as implemented in the Vienna Ab-initio Simulation Package (VASP) [60,61]. Based on the high-throughput calculations for the 1481 non-magnetic NCS insulators that potentially have multiple ST shapes in the BZ, we evaluate the ODPs. We find that there are 64 compounds within ODP1 (i.e., linear-in-$k$ SSs larger than 1 meV at band edges and $k$-valleys with different CPGS categories). The final number of selected compounds depends on the threshold used for $\Delta_{k_1 k_2}$, which in turns depends on the resolution of the measurement and the specific device application. For instance, only 37 compounds have enough small energy difference between states at different valleys (ODP2), when we use $\Delta_{k_1 k_2} = 100$ meV. Tables V and VI show the experimentally synthesized compounds at the convex hull (i.e., energy above the convex hull equal to zero $E_{ch}$=0 meV) obtained in the inverse design process. For each compound we present the ICSD code, energy above the convex hull ($E_{ch}$) given by the materials project (meV/fu), spin splitting (SS) in meV, and R-factor for refinement of the experimental structure. The double entries stand for compounds with the same atomic composition but different ICSD number and different SS.

**B. Discussion of the predicted compounds for potential spin-valeytronics applications**

The list of compounds identified to have multiple ST shapes in the BZ includes known ferroelectrics as GeTe and also 36 previously synthesized compounds potentially allowing the spin-polarization control based on the $k$-valley energy. Table V and VI show the 12 binary and 25 ternary rationally selected compounds. Here, we explore in details some of these compounds.

Our inverse design approach confirms the multiple STs in GeTe (Rashba Hamiltonian at the Z and Rashba-Dresselhaus Hamiltonian at the A point), which is very well known because of its ferroelectric properties [13,19,25] (Table V). We find other fabricated compound with giant SS near band edges. For example, $AsSe_3Tl_3$ and $SrIr_2P_2$ have SS larger than 70 and 80 meV, respectively. While $AsSe_3Tl_3$ has Rashba and Rashba-Dresselhaus STs at the Z and L high-symmetry $k$-points, respectively, in $SrIr_2P_2$ the Weyl ST is found at the $\Gamma$ point and an undefined ST at the M point (Table VI). On the other hand, the DFT calculated Weyl, Rashba and Dresselhaus coefficients ($\lambda_W$, $\lambda_R$, and $\lambda_D$) are usually small (< 1 eV/Å). Indeed, only few compounds have large $\lambda$ coefficient, e.g., $NTlO_2$ with Weyl coefficient $\lambda_W$ of 1.94 eV/Å at the $\Gamma$ point and 1.10 eV/Å at the high-symmetry point H. A remarkable example of a avery well knos compound with relatively weak SOC is the non-polar chiral $SiO_2$ ($P6_522$), CPGS=$D_6$, that have radial ST at the high-symmetry point A (WPGS=CPGS) and tangential-radial ST at the high-symmetry point H (WPGS=$D_3$), as shown in Fig. 9. The SS at these $k$-points are 26 and 6 meV, respectively.

Table V. Binary selected compounds with multiple ST shapes in the BZ. The energy difference between states at different valleys is $\Delta_{k_1 k_2} = 100$ meV. The space group (SG), SG index, CPGS, and high symmetry k-points with non-zero spin splitting (SS) are given for each compound, as well as the ratio between the SS and momentum offset $\lambda = \Delta_{ss}/k$ (eVÅ). The SOC Hamiltonian predicted by the WPGS (Fig. 3) is given for each high-symmetry k-point.

| Material | ICSD | SG index | CPGS | k-point | WPGS | SOC Hamiltonian | SS (meV) | $\lambda$ (eVÅ) |
|---|---|---|---|---|---|---|---|---|
| Bi$_2$O$_3$ | 168810 | R3m (160) | $C_{3v}$ | F | $C_{3v}$ | R | 7 | 0.20 |
| | | | | L | $C_s$ | RD | 1 | 0.06 |
| GeTe | 56040 | R3m (160) | $C_{3v}$ | L | $C_s$ | RD | 28 | 3.00 |
| | | | | Z | $C_{3v}$ | R | 209 | 4.25 |
| PbS | 183243 | R3m (160) | $C_{3v}$ | L | $C_s$ | RD | 20 | 2.12 |
| | | | | Z | $C_{3v}$ | R | 56 | 1.95 |
| BaTe$_3$ | 36366 | P$\bar{4}2_1$m (113) | $D_{2d}$ | X | $C_{2v}$ | RD | 11 | 1.09 |
| | | | | G | $D_{2d}$ | D | 2 | 0.26 |
| | | | | Z | $D_{2d}$ | D | 23 | 0.54 |
| Bi$_2$O$_3$ | 41764 | P$\bar{4}2_1$c (114) | $D_{2d}$ | X | $D_2$ | WD | 16 | 0.40 |
| | | | | G | $D_{2d}$ | D | 1 | 0.09 |
| | | | | R | $D_2$ | WD | 13 | 0.36 |
| Bi$_2$O$_3$ | 168808 | P$\bar{4}$m2 (115) | $D_{2d}$ | X | $D_{2d}$ | D | 46 | 0.37 |
| | | | | R | $D_2$ | WD | 36 | 0.32 |
| B$_2$O$_3$ | 16021 | P3$_1$21 (152) | $D_3$ | A | $D_3$ | W | 13 | 0.35 |
| | | | | L | $C_2$ | Undefined | 21 | 0.22 |
| SiO$_2$ | 170542 | P6$_5$22 (179) | $D_6$ | M | $D_2$ | WD | 3 | 0.11 |
| | | | | A | $D_6$ | W | 26 | 0.46 |
| | | | | H | $D_3$ | D | 6 | 0.12 |
| BeF$_2$ | 9481 | P6$_2$22 (180) | $D_6$ | K | $D_3$ | D | 1 | 0.03 |
| | | | | A | $D_6$ | D | 6 | 0.11 |
| | | | | L | $D_2$ | WD | 1 | 0.05 |
| SnI$_4$ | 18010 | P$\bar{4}$3m (215) | $T_d$ | X | $D_{2d}$ | D | 19 | 0.64 |
| | | | | M | $D_{2d}$ | D | 9 | 0.21 |
| | | | | G | $T_d$ | D | 7 | 0.10 |
| | | | | R | $T_d$ | D | 4 | 0.12 |
| SiO$_2$ | 75647 | P$\bar{4}$3m (215) | $T_d$ | X | $D_{2d}$ | D | 2 | 0.31 |
| | | | | M | $D_{2d}$ | D | 5 | 0.12 |
| H$_2$O$_2$ | 34253 | P4$_1$2$_1$2 (92) | $D_4$ | X | $D_2$ | WD | 22 | 0.32 |
| | | | | Z | $D_4$ | W | 4 | 0.37 |

Table V. Ternary selected compounds with multiple ST shapes in the BZ. The energy difference between states at different valleys is $\Delta_{k_1 k_2} = 100$ meV. The space group (SG), SG index, CPGS, and high symmetry k-points with non-zero spin splitting (SS) are given for each compound, as well as the ratio between the SS and momentum offset $\lambda = \Delta_{ss}/k$ (eVÅ). The SOC Hamiltonian predicted by the WPGS (Fig. 3) is given for each high-symmetry k-point.

| Material | ICSD | SG index | CPGS | k-point | WPGS | SOC Hamiltonian | SS (meV) | $\alpha_R$ (eVÅ) |
|---|---|---|---|---|---|---|---|---|
| Nb$_3$Sb$_2$Te$_5$ | 417101 | I$\bar{4}$3m (217) | $T_d$ | H | $T_d$ | D | 6 | 0.23 |
| | | | | N | $C_{2v}$ | RD | 24 | 0.86 |
| AsSe$_3$Tl$_3$ | 15148 | R3m (160) | $C_{3v}$ | L | $C_s$ | RD | 77 | 1.22 |
| | | | | Z | $C_{3v}$ | R | 12 | 0.53 |
| LaTlO$_3$ | 200088 | P6$_3$mc (186) | $C_{6v}$ | M | $C_{2v}$ | RD | 8 | 1.06 |
| | | | | G | $C_{6v}$ | R | 8 | 0.56 |
| Ag$_5$SbS$_4$ | 16987 | Cmc2$_1$ (36) | $C_{2v}$ | S | $C_2$ | Undefined | 5 | 0.67 |
| | | | | G | $C_{2v}$ | RD | 3 | 0.41 |
| | | | | Y | $C_{2v}$ | RD | 10 | 0.39 |
| BaZnF$_4$ | 182605 | Cmc2$_1$ (36) | $C_{2v}$ | R | $C_2$ | Undefined | 2 | 0.24 |
| | | | | Y | $C_{2v}$ | RD | 9 | 0.11 |
| BaF$_4$Zn | 402926 | Cmc2$_1$ (36) | $C_{2v}$ | R | $C_2$ | Undefined | 2 | 0.23 |
| | | | | Y | $C_{2v}$ | RD | 10 | 0.13 |
| ClMnO$_3$ | 416749 | Cmc2$_1$ (36) | $C_{2v}$ | G | $C_{2v}$ | RD | 2 | 0.06 |
| | | | | A1 | $C_2$ | Undefined | 2 | 0.06 |
| LaTaO$_4$ | 97688 | Cmc2$_1$ (36) | $C_{2v}$ | R | $C_2$ | Undefined | 40 | 0.48 |
| | | | | G | $C_{2v}$ | RD | 1 | 0.03 |
| | | | | A1 | $C_2$ | Undefined | 31 | 0.37 |
| Na$_2$PtS$_2$ | 87219 | Cmc2$_1$ (36) | $C_{2v}$ | S | $C_2$ | Undefined | 1 | 0.11 |
| | | | | R | $C_2$ | Undefined | 10 | 0.70 |
| | | | | G | $C_{2v}$ | RD | 7 | 0.14 |
| | | | | Y | $C_{2v}$ | RD | 1 | 0.07 |
| Na$_2$PtSe$_2$ | 40429 | Cmc2$_1$ (36) | $C_{2v}$ | S | $C_2$ | Undefined | 1 | 0.09 |
| | | | | R | $C_2$ | Undefined | 9 | 0.85 |
| | | | | Y | $C_{2v}$ | RD | 1 | 0.09 |
| AgC$_2$N$_3$ | 843 | P3$_1$21 (152) | $D_3$ | L | $C_2$ | Undefined | 1 | 0.09 |
| | | | | M | $C_2$ | Undefined | 1 | 0.09 |
| | | | | G | $D_3$ | W | 2 | 0.56 |
| AlAsO$_4$ | 33834 | P3$_1$21 (152) | $D_3$ | A | $D_3$ | W | 2 | 0.61 |
| | | | | L | $C_2$ | Undefined | 22 | 0.15 |
| BF$_4$Li | 171375 | P3$_1$21 (152) | $D_3$ | A | $D_3$ | W | 5 | 0.10 |
| | | | | L | $C_2$ | Undefined | 6 | 0.06 |
| | | | | G | $D_3$ | W | 8 | 0.17 |
| BaZnO$_2$ | 25812 | P3$_1$21 (152) | $D_3$ | M | $C_2$ | Undefined | 1 | 0.05 |
| | | | | A | $D_3$ | W | 4 | 0.34 |
| | | | | G | $D_3$ | W | 9 | 0.19 |
| | | | | L | $C_2$ | Undefined | 13 | 0.79 |
| CsNO$_2$ | 50327 | P3$_1$21 (152) | $D_3$ | L | $C_2$ | Undefined | 1 | 0.10 |
| | | | | G | $D_3$ | W | 5 | 1.02 |
| NTlO$_2$ | 50325 | P3$_1$21 (152) | $D_3$ | L | $C_2$ | Undefined | 1 | 0.13 |
| | | | | H | $D_3$ | W | 6 | 1.10 |
| | | | | G | $D_3$ | W | 10 | 1.94 |
| | | | | A | $D_3$ | W | 80 | 0.83 |
| CaIr$_2$P$_2$ | 95756 | P3$_2$21 (154) | $D_3$ | G | $D_3$ | W | 40 | 0.96 |
| | | | | M | $C_2$ | undefined | 77 | 0.85 |
| SrIr$_2$P$_2$ | 73531 | P3$_2$21 (154) | $D_3$ | G | $D_3$ | W | 53 | 0.96 |
| | | | | M | $C_2$ | undefined | 87 | 0.84 |
| LaPO$_4$ | 31564 | P6$_2$22 (180) | $D_6$ | M | $D_2$ | WD | 3 | 0.49 |
| | | | | A | $D_6$ | W | 4 | 0.14 |
| Hg$_3$Te$_2$Br$_2$ | 27402 | I2$_1$3 (199) | $T$ | N | $C_2$ | undefined | 6 | 0.56 |
| | | | | G | $T$ | W | 10 | 0.27 |
| Cl$_2$Hg$_3$S$_2$ | 28159 | I2$_1$3 (199) | $T$ | G | $T$ | W | 3 | 0.14 |
| | | | | N | $C_2$ | undefined | 4 | 0.17 |
| Cl$_2$Hg$_3$Te$_2$ | 27401 | I2$_1$3 (199) | $T$ | G | $T$ | W | 9 | 0.31 |

| | | | | N | $C_2$ | undefined | 32 | 0.68 |
|---|---|---|---|---|---|---|---|---|
| PtSnTh | 108712 | F$\bar{4}$3m (216) | $T_d$ | G | $T_d$ | D | 4 | 0.33 |
| | | | | X | $D_{2d}$ | D | 6 | 0.76 |
| GaPO$_4$ | 30881 | P3$_1$21 (152) | $D_3$ | G | $D_3$ | W | 2 | 0.54 |
| | | | | L | $C_2$ | undefined | 3 | 0.17 |
| GaPO$_4$ | 33253 | P3$_1$21 (152) | $D_3$ | G | $D_3$ | W | 2 | 0.50 |
| | | | | L | $C_2$ | undefined | 3 | 0.17 |

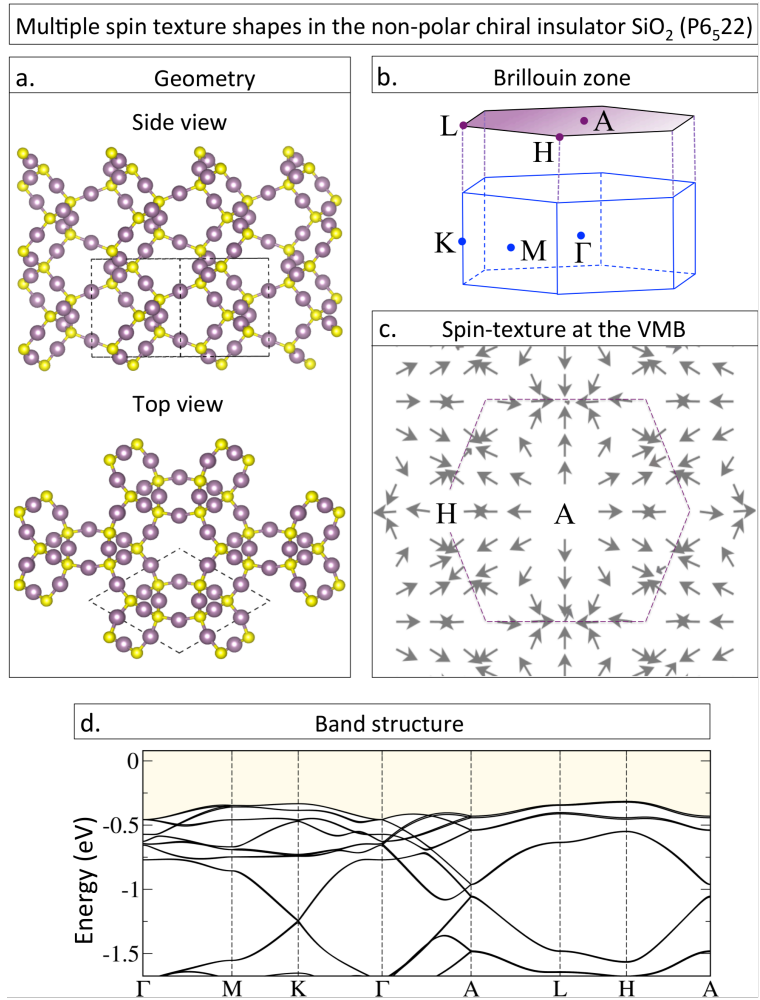

Figure 9. Multiple spin texture shapes in the non-polar chiral insulator SiO$_2$ (P6522). (a) Top and side view of SiO$_2$. Silicon and Oxygen atoms are represented in yellow and purple, respectively. (b) Brillouin zone for the chiral SiO$_2$. The two-dimensional hexagonal projection of the plane containing the high symmetry k-points A and H is shown in purple. (c) Spin-polarization at the valence band maximum (VBM) at the two-dimensional projection in (b). Around A, the radial ST is observed, while the tangential-radial ST is found around H. The hexagonal BZ is delimited by the purple region. The arrows stand for the in-plane spin-polarization components and the out-plane spin-component $S_z$ is always zero. (d) DFT calculated valence bands for SiO$_2$ (P6522).

## VII. Conclusions

Despite the fact that the crystal point group symmetry (CPGS) (e.g., absence or presence of electric dipoles) does not unequivocally determine the spin texture (ST), we find that the wavevector point group symmetry (WPGS) $G^*(k_0)$ can be a descriptor for this functionality. The important consequence of this discovery is that the selectivity in types of STs can be rationalized on the basis of symmetry (not spin-orbit physics neither the mere existence of electric fields/dipoles), and therefore the ST can be *designed.* These consequences are extended to other spin related phenomena. For instance, the *spin-momentum locking* effect is enforced by symmetry, rather than a consequence of strong SOC or topological effects. Additionally, our findings suggest the possibility of accessing different ST in the same compound by controlling the relative energy position of states in different valleys of the Brillouin zone, which has potential application for spin-valleytronics. We use the symmetry conditions defining the ST to establish all possible ST prototypes in the 21 NCS crystal point group symmetry for three-dimensional crystals. Using these symmetry relations as design principles, we select 1481 compounds from the aflow-ICSD database [29]. Performing DFT band structure calculations for the selected compounds, we predict 37 materials unnoticed to have multiple ST shapes at different $k$-valleys. The ST symmetry classification as well as the predicted compounds with multiple ST can be a platform for potential application for the control of the ST by accessing to different valleys.

**Acknowledgments**

The work at the University of Colorado at Boulder was supported by the National Science Foundation NSF Grant NSF-DMR-CMMT No. DMR-1724791. CMA and GMD are supported by the São Paulo Research Foundation (FAPESP) Grants No. 18/11856-7, 19/03663-7 and 17/02317-2. High throughput calculations were performed at the Santos Dumont supercomcomputer (LNCC/Brazil). We thank Linding Yuan for the valuable discussion on the polar symmetry $C_s$ and its spin texture.